\documentclass[iop]{emulateapj}

\slugcomment{To appear in the Astrophysical Journal}

\shorttitle{Binary disks in Taurus}
\shortauthors{Akeson \& Jensen}

\begin{document}


\title{Circumstellar disks around binary stars in Taurus}


\author{R. L. Akeson\altaffilmark{1} and E. L. N. Jensen\altaffilmark{2}}

\altaffiltext{1}{NASA Exoplanet Science Institute, IPAC/Caltech, Pasadena, CA, 91125 \\contact:  rla@ipac.caltech.edu}
\altaffiltext{2}{Swarthmore College, Dept of Physics and Astronomy, Swarthmore PA 19081}

\begin{abstract}

We have conducted a survey of 17 wide ($>$ 100 AU) young binary systems in Taurus with
the Atacama Large Millimeter Array (ALMA) at two wavelengths.  
The observations were designed to measure the masses of circumstellar disks in these
systems as an aid to understanding the role of multiplicity in star and planet formation.
The ALMA observations had sufficient resolution to localize emission within the binary system.  
Disk emission was detected around all primaries and ten secondaries,
with disk masses as low as 10$^{-4}$ M$_{\odot}$.  We compare the properties of our sample to the population
of known disks in Taurus and find that the disks from this binary sample match the 
scaling between stellar mass and millimeter flux of F$_{mm} \propto$ M$_{\ast}^{1.5-2.0}$ to within the scatter found in
previous studies. 
We also compare the properties of the primaries to those of the
secondaries and find that the secondary/primary stellar and disk mass \emph{ratios} are not correlated; in three systems, the circumsecondary disk is more massive
than the circumprimary disk, counter to some theoretical predictions.


\end{abstract}

\section{Introduction}
\label{intro}

Most stars are formed in binary or multiple systems and remain in such systems for their main 
sequence lifetimes \citep [e.g.][]{mon07}. Therefore, understanding the causes and effects 
of multiplicity is an essential ingredient of complete models of both star and planet formation.   
Circumstellar disks play a crucial role in both processes, tracing effects of different 
binary formation mechanisms, providing conduits for material to accrete onto the stars, 
and serving as the reservoir of raw material for planet formation.   

At a given point in time, the distribution of observed disk masses is a function
of the initial disk masses and disk evolution.  For multiple
systems, dynamical interactions between the stars, the circumstellar disks,
and any circumbinary material will also impact both the disk formation and evolution.
Models of binary star formation by \citet{bat00} predict that the
circumprimary disk, i.e. the disk around the more massive star, will
have more mass than the circumsecondary disk; however, these
models do not follow the viscous evolution of the disk after the formation stage.
Observations to date largely
support the prediction of a more massive circumprimary disk,  although the sample of systems observed is relatively small 
and generally comprise only the brightest sources. \citet{jen03} found that the primary 
star had the most massive disk in all four young binaries they observed; 
indeed in only one system was the secondary's disk detected at all, despite most of the 
secondaries showing signs of accretion. More recent work by \citet{har12}
has expanded the number of observed binary systems and also found that when both components were detected, the primary had higher flux, but with sensitivity levels of a few mJy, 
many secondaries remained undetected.

Planet formation in these systems may also be impacted as models of the interactions of binary 
stars with their associated disks predict that the disks will be truncated somewhere 
between 0.2 and 0.5 times the binary separation, depending on the eccentricity of the 
system \citep{art96}.   However, these models do not
address the surface density and evolution of the remaining disk material.
If the secondary disks retain roughly the same surface density as
the inner parts of disks around single stars, then they may still retain
enough mass to form planets.  Previous observations 
have not had the sensitivity to distinguish between disks that are simply truncated, 
and those that have been significantly depleted by further accretion.   
Disk models show that truncation effects can affect the observed flux for separations up to a 
few hundred AU \citep{jen96}.

The essential question for planet formation, then, is whether or not the disks 
around individual components of close binary stars are similar to the inner regions 
of disks around single stars. Early observations demonstrated that the 
unresolved millimeter emission, which traces the dust in the outer regions 
of the disk, is indeed reduced, consistent with truncation \citep{bec90,ost95,jen96}. 
But most observations of binaries with separations in the ranges of 50-100 AU have 
yielded upper limits rather than detections, and indeed only about half of all young 
binaries in Taurus have been detected at all at millimeter wavelengths, despite the 
fact that many more than half of them were detected by IRAS at 60 $\mu$m.
With the advent of ALMA observations, which provide a substantial increase
in sensitivity at the required resolution,
it is now possible to reach much 
lower disk surface densities, and possibly to detect very low mass protoplanetary disks.

To address these issues, we have obtained ALMA Cycle 0 observations of 17 young binary systems in Taurus for which the components
can be resolved.  In \S \ref{sample} we describe the sample selection and properties, in \S \ref{obs}
we describe the ALMA observations and data reduction, in \S \ref{results} we present the
results,  and we give our conclusions in \S \ref{conclusions}.

\begin{deluxetable*}{llllll}[b!]
\tabletypesize{\footnotesize}
\tablewidth{0pt}
\tablecaption{Taurus Binary Sample}
\tablehead{
\colhead{Source Name}           & \colhead{Additional names}      &
\colhead{Separation}  & 
\colhead{mm flux\tablenotemark{a}}         
 & \colhead{Spectral types\tablenotemark{b}} 
 & \colhead{Previous mm} \\
& &  \colhead{(AU)}  & 
\colhead{(mJy)} & & \colhead{detection\tablenotemark{c}} 
}
\startdata
FV Tau & & 101 & 48$\pm$5 & K5/K6 & 1,2\\
HBC 387 & FV Tau/c & 104 & $<$25 & M2.5/M3.5 & --\\
FQ Tau & & 110 & 28$\pm$7 & M3/M3.5 & 1\\
UY Aur & & 120 & 102$\pm$6 & M0/M2.5 & 1,2\\
FX Tau & & 130 & 17$\pm$3 & M1/M4 & 1\\
HBC 411 & CoKu Tau/3 & 290 & $<$8 & M1/M4.5 & --\\
IRAS 05022+2527 & CIDA 9 & 320 & 71$\pm$7 & K8/M1.5 & 1\\
HK Tau & & 340 & 130$\pm$2 & M0.5/M2 & 1,2\\
IT Tau & & 340 & 22$\pm$3 & K3/M4 & 1,2\\
DK Tau & & 350 & 80$\pm$10 & K8/M1 & 1\\
GK Tau & & 340 & 33$\pm$7 & K7 & 1\\
HN Tau & & 430 & 29$\pm$3 & K5/M4 & 1 \\
V710 Tau & & 450 & 152$\pm$6 & M0.5 & 1\\
IRAS 04113+2758 & MHO 1/2& 550 & 380$\pm$3  & M2.5/M2.5 & 1,2\\
IRAS 04298+2246 & JH 112 & 920 & 30$\pm$10 & K6/M8.5 & 1\\
HO Tau & & 970 & 44$\pm$6 & M0.5 & 1 \\
DS Tau & & 990 & 39$\pm$4 & K5 & 1\\
\enddata
\label{tab:sample}
\tablenotetext{a}{Millimeter fluxes are single-dish fluxes at 850 $\mu$m taken from \citet{and05} except for HK Tau \citep[850 $\mu$m, interferometry]{har12} and IRAS 04113+2758 \citep[1.3 mm, interferometry]{har12}  }
\tablenotetext{b}{Spectral types from \citet{and13}}
\tablenotetext{c}{1 = primary, 2 = secondary}
\end{deluxetable*}

\section {Sample}
\label{sample}

We selected 
targets from a single star formation region, Taurus (distance $\sim$ 140 pc), so that 
effects such as age and cluster environment are kept constant as much as 
possible.   Taurus is ideal in having a significant population of YSOs that 
have evolved into the disk-only state (with no remaining envelope) and is 
very well studied, containing both a well known set of single stars with 
disks for comparison and a significant population of binaries where both 
stellar components have been characterized in the optical or near-infrared.  
We started with the list of known Taurus binaries \citep{and05,kra09} and selected those 
with separations in the range of 0\farcs7 (100 AU) to 10\arcsec\ (1400 AU).    
The inner cutoff was selected such that the two components could be clearly resolved 
with the resolution offered in Cycle 0, while the outer cutoff was chosen to ensure that the systems are likely 
to be physically associated.    We 
eliminated systems classified as Class I from their spectral energy distributions, 
as these systems often contain substantial envelope emission that must be disentangled 
from the disk emission, and we eliminated systems with no active accretion signatures 
that had not been previously detected at millimeter wavelengths.  The resulting 
sample contains 17 systems (Table \ref{tab:sample}) and includes all
Class II Taurus binaries with separations of 100 to 1400 AU from \citet{and05} and \citet{kra09}. 
Higher-order multiple systems were excluded where known at the time of our
sample selection, although
two were observed (see notes in \S \ref{individ}).

\section{Observations}
\label{obs}

The observations dates and
ALMA data set names are given in Table \ref{tab:obs}. 
The correlator was configured with each of the four basebands covering a total bandwidth of 1.875 GHz
with a channel spacing of 488 kHz.  At 1.3 mm (Band 6), one of the correlator basebands was set to cover
the CO(2-1) transition at 230.5 GHz, while at 850 $\mu$m (Band 7), one baseband covered CO(3-2) at 345.8 GHz.
Each target source was observed only once per band and was bracketed by observations
of the gain calibrator, J051002+180041.   Data for each band were calibrated separately
using the CASA package and scripts provided by the NRAO ALMA center.  The
system temperature, water vapor phase corrections, and flagging were applied using the
standard scripts.  The amplitude and phase of the passband were calibrated against
J0423-013.  The absolute flux calibration used Callisto and the Butler-JPL-Horizons  2012 flux models, which
resulted in a zero spacing flux of 8.54 Jy at 1.3 mm and 19.45 Jy at 850 $\mu$m.
The data taken in April 2012 at 1.3 mm showed a much lower gain stability than the
other 1.3 mm data sets and are not used here.

\begin{deluxetable*}{llll}[tbp]
\tabletypesize{\scriptsize}
\tablewidth{0pt}
\tablecaption{Observation Log \label{tab:obs}
}
\tablehead{
\colhead{ALMA band/}     &\colhead{Data set}      & \colhead{Observation}      &
\colhead{Sources} \\
\colhead{wavelength}& & \colhead{Date (UT)} \\
}
\startdata
Band 7/ & 2011.0.00150.S\_2012-12-05 & 11/16/2012 & FV Tau, HBC 387, FQ Tau, FX Tau, HK Tau, DK Tau, IRAS 04113+2758  \\
850 $\mu$m& 2011.0.00150.S\_2012-12-06 & 11/16/2012  & HBC 411, IT Tau, GK Tau, HN Tau, V710 Tau, IRAS 04298+2246, HO Tau \\
& 2011.0.00150.S\_2012-12-07 & 11/16/2012  & UY Aur, IRAS 05022+2527, DS Tau \\
Band 6/ & 2011.0.00150.S\_2012-12-12 & 11/17/2012  & HBC 411, IT Tau, GK Tau, HN Tau, V710 Tau, IRAS 04298+2246, HO Tau \\
1.3 mm & 2011.0.00150.S\_2012-12-19 & 11/17/2012 & FV Tau, HBC 387, FQ Tau, FX Tau, HK Tau, DK Tau, IRAS 04113+2758  \\
& 2011.0.00150.S\_2012-12-20 & 11/17/2012  & UY Aur, IRAS 05022+2527, DS Tau \\
& 2011.0.00150.S\_2013-01-26 & 4/22/2012 & FV Tau, HBC 387, FQ Tau, FX Tau, HBC 411 \\
\enddata
\end{deluxetable*}

Continuum and CO images at each band were generated using the \emph{clean} task within CASA, with
a robust beam weighting of 0.5.  
Each data set had at least one target with sufficient
continuum flux to allow phase-only self-calibration, which was applied after
the other calibrations.   For the 2012-12-05 and 2012-12-19 data sets, the
self-calibration reference source was IRAS 04113+2758, except
for HK Tau and DK Tau, which were used as their own reference.  For the 2012-12-06 and 2012-12-12 data sets, the
self-calibration reference source was V710 Tau and for 2012-12-07 and
2012-12-20, each source was used as its own self-calibration reference.
Given the short time on source (60 sec at 1.3 mm and 90 sec at 850 $\mu$m), the continuum
data were time-averaged to a single point when calculating the self-calibration corrections.
After the phase self-calibration was applied,  images were generated
interactively with 50 iterations per cycle and clean boxes placed only around
emission visible in the dirty map.  The cycles were stopped when the
residuals in the clean boxes were at or below the rms in the rest of the image.
Most sources required 2 cycles, while the brightest sources required 3 or 4.
The primary beam correction was applied and the entire primary
beam was mapped for each source; no continuum emission was detected away
from the known source positions.
FV Tau and HBC 387 (FV Tau/c) are separated by 12\arcsec, but were observed
in two separate pointings. As HBC 387 was only marginally detected in its single pointing, we combined
the two observations in a mosaic, increasing the sensitivity.

\begin{figure*}[h]
\includegraphics[height=8in]{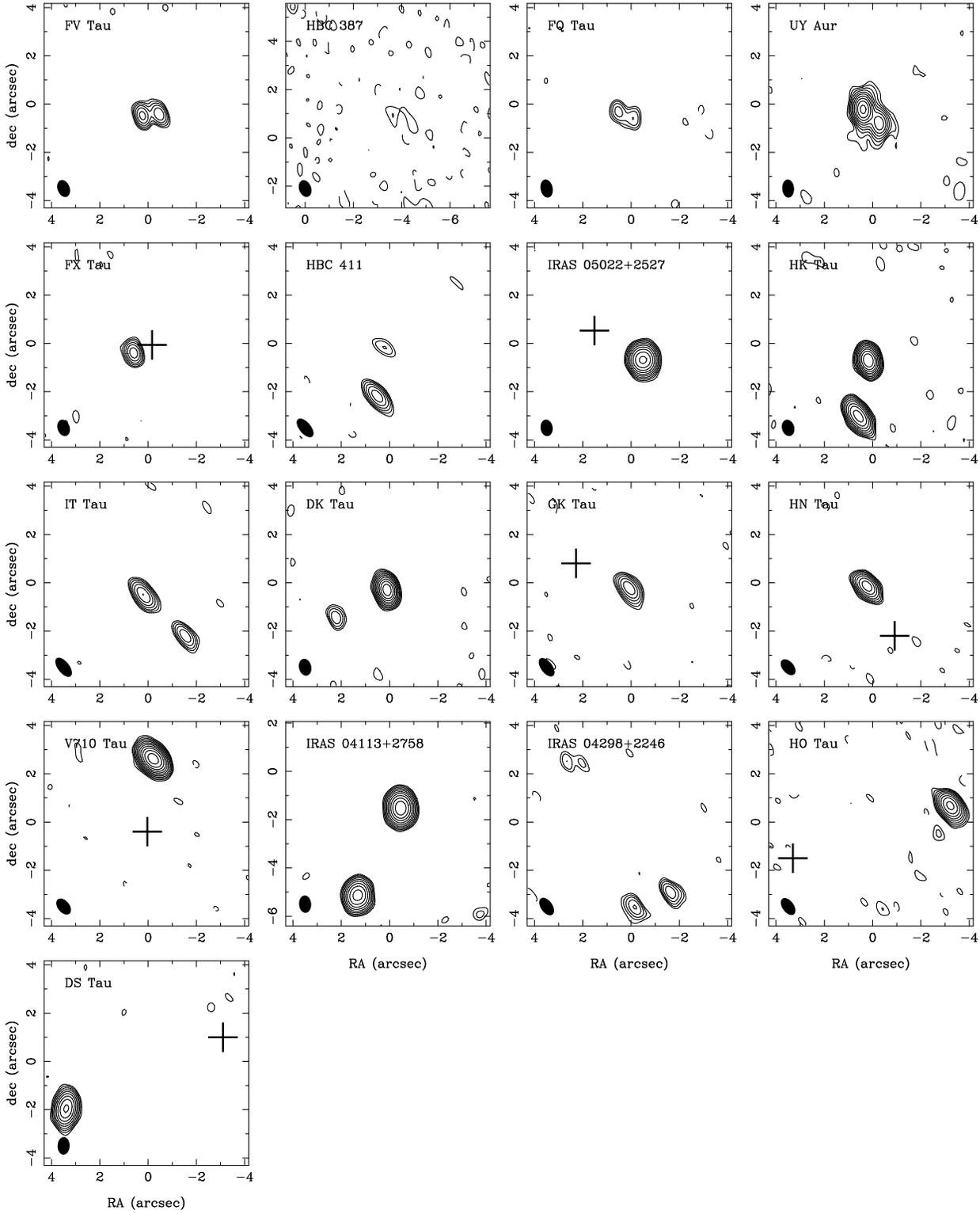}
\caption{850 $\mu$m cleaned continuum images for all sources.  Contour levels start at 3$\sigma$ and increase by 50\% in each step,
except for HBC 387 where the levels are 2 and 4$\sigma$.  Negative contours are shown with dashed lines.  The clean beam is shown in the bottom left for each source.  A plus sign shows the stellar positions for undetected secondaries.
\label{fig:images}
}
\end{figure*}

The 850 $\mu$m continuum maps are shown in Figure \ref{fig:images}.
The images use the default restoring beam, which is
a Gaussian fit to the dirty beam.  These beam sizes are listed in Table \ref{tab:fit}.
In every system, the primary component has been detected.  The two wide
components of IRAS 04298+2246 (JH 112) are each resolved into two close components
and we treat this system as two separate binaries in \S \ref{results}.  
For systems where the secondary is not detected,
its position is marked with a plus.  The typical rms values achieved are
0.15-0.20 mJy/beam at 1.3 mm and 0.35-0.40 mJy/beam at 850 $\mu$m,
significantly more sensitive than previous surveys in Taurus.  The images
of the brightest source, IRAS 04113+2758, are dynamic range limited
and have higher rms levels.  

Examination of the CO images show that CO is detected for all sources except 
HBC 387 (CoKu Tau/3), FQ Tau, and IRAS 04298+2246.   As seen
in Table \ref{tab:fit}, these are the three weakest sources in continuum emission.
In the detected sources, the integrated CO emission is not correlated
with continuum flux but we also note that the cloud contamination
varies considerably from source to source.

\section{Results}
\label{results}

The observational design of this program was to detect emission from
circumstellar disks but not to characterize the physical parameters of
the disks.  Fitting detailed disk models to determine temperature, density,
and other physical parameters is best done in the uv-plane \citep[see the discussion in][]{dut07}.   As our
goal was to compare overall properties of the disks (flux and mass) and
because CASA does not currently allow multiple component fitting
in the uv-plane, we fit these data in the image plane to
determine the total flux and size.  
The clean continuum maps at each band were fit with two-dimensional Gaussians for each source using the CASA
routine \emph{imfit}.  The fitting results are presented in Tables \ref{tab:fit} and \ref{tab:pos}.  
The positions and binary separations were measured from the 850 $\mu$m images before
self-calibration while the total flux and deconvolved size were determined from fits
to the self-calibrated maps.
If the uncertainty on the major axis was larger than the fit size, we list that object as a
point source.  For the resolved disks, we include the position angle from the Gaussian
fit and the derived inclination angle.  Upper limits given on the flux are 3$\sigma$.

\begin{deluxetable*}{llllllll}[h]
\tabletypesize{\tiny}
\tablewidth{0pt}
\
\tablecaption{Gaussian Fitting Results \label{tab:fit}}
\tablehead{
\colhead{Source}   
& \colhead{Peak flux} & \colhead{$\sigma$ } & \colhead{Beam size} & \colhead{Total flux} & \colhead{Deconvolved size} & \colhead{Position angle}& \colhead{Inclination} \\ 
&  \colhead{(mJy)} & \colhead{(mJy)}       & \colhead{(arcsec)} & \colhead{(mJy)} & \colhead{(mas)} & \colhead{(deg)}& \colhead{(deg)} \\
}
\startdata
&\multicolumn{7}{c}{1.3 mm} \\ 
\tableline

FV Tau A&    7.0 &  0.15 & 1.12x0.74 &   6.17$\pm$0.16  &       point source  &          --- &     --- \\ 
FV Tau B&    6.0 &  0.15 & 1.12x0.74 &   5.93$\pm$0.18 &   point source  &          --- &   ---    \\ 
HBC 387 A  &    0.5 &  0.12 & 1.10x0.73 &   0.76$\pm$0.17  &       point source  &          ---  &     --- \\ 
HBC 387  B&    0.5 &  0.12 & 1.10x0.73 &   0.40$\pm$0.07 &       point source  &        ---   &   ---    \\ 
FQ Tau A&    2.7 &  0.15 & 1.14x0.72 &   3.10$\pm$0.21  &   point source  &          ---   &     --- \\ 
FQ Tau B&    2.5 &  0.15 & 1.14x0.72 &   1.63$\pm$0.13 &       point source  &        ---   &   ---    \\ 
FX Tau A&    7.1 &  0.15 & 1.05x0.73 &   7.03$\pm$0.15  &       point source  &          ---  &     --- \\ 
FX Tau B& $<$0.45 &  0.15 & 1.05x0.73 & not detected &              ---    &        ---   &   ---    \\ 
HK Tau A&   33.0 &  0.15 & 1.06x0.73 &  33.80$\pm$0.20  &  point source  &          --- &     --- \\ 
HK Tau B&   12.6 &  0.15 & 1.06x0.73 &  16.01$\pm$0.24 &       point source  &        ---   &   ---    \\ 
DK Tau A&   29.7 &  0.14 & 1.09x0.72 &  30.30$\pm$0.18  &       point source  &          ---  &     --- \\ 
DK Tau B&    2.7 &  0.14 & 1.09x0.72 &   2.88$\pm$0.19 &  365$\pm$ 46 x  120$\pm$ 70  &  26.6$\pm$  1.2 & 70$\pm$13  \\ 
IRAS 04113+2758 A&  179.0 &  0.54 & 1.15x0.73 & 216.74$\pm$0.76  &  403$\pm$  8 x  397$\pm$  8  & 175.0$\pm$ 94.0  & 10$\pm$ 2 \\ 
IRAS 04113+2758 B&  106.0 &  0.54 & 1.15x0.73 & 133.34$\pm$0.79 &  472$\pm$ 14 x  384$\pm$ 17  & 126.4$\pm$  8.5 & 35$\pm$ 3  \\ 
UY Aur A&   19.9 &  0.20 & 1.24x0.74 &  20.83$\pm$0.23  &  336$\pm$ 33 x  184$\pm$ 73  & 177.0$\pm$ 11.0  & 56$\pm$16 \\ 
UY Aur B&    8.4 &  0.20 & 1.24x0.74 &   7.87$\pm$0.25 &   point source  &          ---  &   ---    \\ 
IRAS 05022+2527 A&   28.7 &  0.19 & 1.09x0.76 &  35.22$\pm$0.26  &  473$\pm$ 16 x  324$\pm$ 23  & 106.1$\pm$  6.4  & 46$\pm$ 4 \\ 
IRAS 05022+2527 B& $<$0.57 &  0.19 & 1.09x0.76 & not detected &              ---    &        ---   &   ---    \\ 
DS Tau A&   17.1 &  0.18 & 1.21x0.76 &  19.94$\pm$0.25  &  581$\pm$ 23 x  203$\pm$ 76  & 154.1$\pm$  3.0  & 69$\pm$ 8 \\ 
DS Tau B& $<$0.54 &  0.18 & 1.21x0.76 & not detected &              ---    &        ---   &   ---    \\ 
HBC 411 A&    1.8 &  0.27 & 1.23x0.74 &   1.84$\pm$0.30  &       point source  &          ---  &     --- \\ 
HBC 411 B&    5.8 &  0.27 & 1.23x0.74 &   5.79$\pm$0.29 &       point source  &        ---   &   ---    \\ 
IT Tau A&    7.1 &  0.21 & 1.26x0.74 &   7.00$\pm$0.24  &       point source  &          ---  &     --- \\ 
IT Tau B&    4.0 &  0.21 & 1.26x0.74 &   4.17$\pm$0.27 &       point source  &        ---   &   ---    \\ 
GK Tau A&    5.2 &  0.22 & 1.20x0.74 &   5.33$\pm$0.56  &  205$\pm$ 77 x   58$\pm$126  &  93.1$\pm$  3.7  & 73$\pm$59 \\ 
GK Tau B& $<$0.66 &  0.22 & 1.20x0.74 & not detected &              ---    &        ---   &   ---    \\ 
HN Tau A&   12.8 &  0.20 & 1.05x0.74 &  13.46$\pm$0.26  &       point source  &          ---  &     --- \\ 
HN Tau B& $<$0.60 &  0.20 & 1.05x0.74 & not detected &              ---    &        ---   &   ---    \\ 
V710 Tau A&   52.4 &  0.21 & 1.06x0.74 &  59.18$\pm$0.33  &  343$\pm$ 16 x  264$\pm$ 22  &  77.0$\pm$ 12.0  & 39$\pm$ 5 \\ 
V710 Tau B& $<$0.63 &  0.21 & 1.06x0.74 & not detected &              ---    &        ---   &   ---    \\ 
IRAS 04298+2246 Aa &    3.7 &  0.19 & 1.14x0.75 &   3.53$\pm$0.23  &       point source  &          ---  &     --- \\ 
IRAS 04298+2246 Ab &    3.1 &  0.19 & 1.14x0.75 &   3.06$\pm$0.24 &       point source  &        ---   &   ---    \\ 
IRAS 04298+2246 Ba &    0.4 &  0.19 & 1.14x0.75 &   0.25$\pm$0.08 &       point source  &        ---   &   ---    \\ 
IRAS 04298+2246 Bb &    1.6 &  0.19 & 1.14x0.75 &   2.30$\pm$0.24  &  655$\pm$210 x  464$\pm$438  & 124.0$\pm$ 71.0  & 44$\pm$115 \\ 
HO Tau A&   16.2 &  0.21 & 1.13x0.75 &  17.06$\pm$0.27  &       point source  &          ---  &     --- \\ 
HO Tau B& $<$0.63 &  0.21 & 1.13x0.75 & not detected &              ---    &        ---   &   ---    \\ 
\tableline
 &\multicolumn{7}{c}{850 $\mu$m}   \\
\tableline
FV Tau A &   14.4 &  0.53 & 0.73x0.50 &  13.76$\pm$0.52  &       point source  &          ---  &     --- \\ 
FV Tau B&   11.8 &  0.53 & 0.73x0.50 &  12.10$\pm$0.56 &       point source  &        ---   &   ---    \\ 
HBC 387 A&    1.3 &  0.31 & 0.72x0.50 &   1.38$\pm$0.40  &       point source  &          ---  &     --- \\ 
HBC 387 B&    1.2 &  0.31 & 0.72x0.50 &   0.94$\pm$0.40 &       point source  &        ---   &   ---    \\ 
FQ Tau A&    5.3 &  0.42 & 0.76x0.50 &   5.42$\pm$0.48  &       point source  &          ---  &     --- \\ 
FQ Tau B&    4.3 &  0.42 & 0.76x0.50 &   4.24$\pm$0.47 &       point source  &        ---   &   ---    \\ 
FX Tau A&   14.7 &  0.49 & 0.70x0.50 &  15.65$\pm$0.54  &   point source  &          ---   &     --- \\ 
FX Tau B& $<$1.47 &  0.49 & 0.70x0.50 & not detected &              ---    &        ---   &   ---    \\ 
HK Tau A&   79.3 &  0.40 & 0.69x0.51 &  73.60$\pm$1.30  &  188$\pm$ 35 x  130$\pm$ 62  &  11.0$\pm$ 85.0  & 46$\pm$29 \\ 
HK Tau B&   41.1 &  0.40 & 0.69x0.51 &  56.50$\pm$1.80 &       point source  &        ---   &   ---    \\ 
DK Tau A&   70.1 &  0.40 & 0.71x0.50 &  67.25$\pm$0.47  &  165$\pm$ 15 x  123$\pm$ 22  &  15.0$\pm$ 18.0  & 41$\pm$11 \\ 
DK Tau B&    6.5 &  0.40 & 0.71x0.50 &   5.81$\pm$0.45 &       point source  &        ---   &   ---    \\ 
IRAS 04113+2758 A&  317.0 &  1.95 & 0.72x0.50 & 476.70$\pm$3.20  &  409$\pm$  7 x  397$\pm$  7  & 112.0$\pm$ 48.0  & 13$\pm$ 2 \\ 
IRAS 04113+2758 B&  185.0 &  1.95 & 0.72x0.50 & 295.20$\pm$3.30 &  488$\pm$ 12 x  344$\pm$ 16  & 117.2$\pm$  4.7 & 45$\pm$ 2  \\ 
UY Aur A&   43.7 &  0.37 & 0.75x0.49 &  48.40$\pm$0.66  &  270$\pm$ 21 x  170$\pm$ 36  & 176.5$\pm$  9.2  & 50$\pm$10 \\ 
UY Aur B&   15.7 &  0.37 & 0.75x0.49 &  18.27$\pm$0.72 &  366$\pm$ 49 x  228$\pm$ 91  & 174.0$\pm$ 26.0 & 51$\pm$20  \\ 
IRAS 05022+2527 A&   50.1 &  0.33 & 0.69x0.50 &  74.40$\pm$2.00  &  481$\pm$ 28 x  352$\pm$ 32  &  99.9$\pm$  6.9  & 42$\pm$ 6 \\ 
IRAS 05022+2527 B& $<$0.99 &  0.33 & 0.69x0.50 & not detected &              ---    &        ---   &   ---    \\ 
DS Tau A&   32.3 &  0.39 & 0.73x0.50 &  41.30$\pm$1.80  &  614$\pm$ 37 x  251$\pm$ 95  & 164.6$\pm$  4.3  & 65$\pm$10 \\ 
DS Tau B& $<$1.17 &  0.39 & 0.73x0.50 & not detected &              ---    &        ---   &   ---    \\ 
HBC 411 A&    3.0 &  0.44 & 0.93x0.47 &   3.32$\pm$0.55  &       point source  &          ---  &     --- \\ 
HBC 411 B&   13.2 &  0.44 & 0.93x0.47 &  14.29$\pm$0.56 &   point source  &          ---  &   ---    \\ 
IT Tau A&   15.2 &  0.44 & 0.94x0.47 &  15.82$\pm$0.56  &  120$\pm$ 17 x   88$\pm$ 33  & 105.8$\pm$  1.8  & 42$\pm$23 \\ 
IT Tau B&    9.2 &  0.44 & 0.94x0.47 &   8.79$\pm$0.52 &       point source  &        ---   &   ---    \\ 
GK Tau A&   13.1 &  0.41 & 0.89x0.47 &  14.72$\pm$0.60  &   point source  &          ---   &     --- \\ 
GK Tau B& $<$1.23 &  0.41 & 0.89x0.47 & not detected &              ---    &        ---   &   ---    \\ 
HN Tau A&   31.3 &  0.47 & 0.77x0.47 &  34.32$\pm$0.60  &  225$\pm$ 41 x  136$\pm$113  &  65.0$\pm$154.0  & 52$\pm$44 \\ 
HN Tau B& $<$1.41 &  0.47 & 0.77x0.47 & not detected &              ---    &        ---   &   ---    \\ 
V710 Tau A&  109.0 &  0.49 & 0.76x0.48 & 143.50$\pm$0.79  &  368$\pm$  8 x  262$\pm$ 11  &  81.7$\pm$  4.1  & 44$\pm$ 2 \\ 
V710 Tau B& $<$1.47 &  0.49 & 0.76x0.48 & not detected &              ---    &        ---   &   ---    \\ 
IRAS 04298+2246 Aa &    8.7 &  0.41 & 0.82x0.48 &   8.85$\pm$0.49  &       point source  &          ---  &     --- \\ 
IRAS 04298+2246 Ab &    6.5 &  0.41 & 0.82x0.48 &   7.75$\pm$0.61 &   point source  &          ---  &   ---    \\ 
IRAS 04298+2246 Ba &    2.2 &  0.41 & 0.82x0.48 &   3.10$\pm$1.90 &       point source  &        ---   &   ---    \\ 
IRAS 04298+2246 Bb &    2.8 &  0.41 & 0.82x0.48 &   3.53$\pm$0.60  &       point source  &          ---  &     --- \\ 
HO Tau A&   33.1 &  0.45 & 0.80x0.47 &  36.50$\pm$1.10  &  point source  &          ---  &     --- \\ 
HO Tau B& $<$1.35 &  0.45 & 0.80x0.47 & not detected &              ---    &        ---   &   ---    \\ 
\enddata
\end{deluxetable*}

\begin{deluxetable*}{lcccccc}[h]
\tabletypesize{\tiny}
\tablewidth{0pt}
\
\tablecaption{Positional Fitting Results \label{tab:pos}}
\tablehead{
\colhead{Binary System}     & \colhead{Primary RA} & \colhead{Primary dec} & \colhead{Secondary RA} & \colhead{Secondary dec} & \colhead{Separation} & \colhead{Position angle} \\
& (J2000) & (J2000) & (J2000) & (J2000) & (arcsec) & (deg) \\
}
\startdata
FV Tau &  4:26:53.550 $\pm$ 0.019 & 26:06:53.903 $\pm$ 0.025 &  4:26:53.498 $\pm$ 0.021 & 26:06:53.956 $\pm$ 0.027 &  0.696 $\pm$ 0.046 &  274.4 $\pm$   3.0 \\ 
HBC 387 (FV Tau/c) &  4:26:54.142 $\pm$ 0.070 & 26:06:51.947 $\pm$ 0.096 &  4:26:54.070 $\pm$ 0.104 & 26:06:51.504 $\pm$ 0.143 &  1.534 $\pm$ 0.303 &  245.5 $\pm$   8.7 \\ 
FQ Tau &  4:19:12.807 $\pm$ 0.045 & 28:29:32.512 $\pm$ 0.065 &  4:19:12.853 $\pm$ 0.042 & 28:29:32.819 $\pm$ 0.062 &  0.682 $\pm$ 0.109 &   63.1 $\pm$   7.1 \\ 
FX Tau &  4:30:29.659 $\pm$ 0.015 & 24:26:44.667 $\pm$ 0.019 &   ---   &   ---   &   ---   &   ---   \\ 
HK Tau &  4:31:50.580 $\pm$ 0.011 & 24:24:17.378 $\pm$ 0.014 &  4:31:50.610 $\pm$ 0.019 & 24:24:15.065 $\pm$ 0.025 &  2.349 $\pm$ 0.036 &  170.1 $\pm$   0.5 \\ 
DK Tau &  4:30:44.252 $\pm$ 0.010 & 26:01:24.506 $\pm$ 0.014 &  4:30:44.407 $\pm$ 0.019 & 26:01:23.363 $\pm$ 0.027 &  3.404 $\pm$ 0.047 &  118.3 $\pm$   0.6 \\ 
IRAS 04113+2758 & 4:14:26.411 $\pm$ 0.031 & 28:05:59.377 $\pm$ 0.045 &   4:14:26.276 $\pm$ 0.018 & 28:06:02.967 $\pm$ 0.026  &  4.010 $\pm$ 0.063 &  333.5 $\pm$   0.6 \\ 
HBC 411 &  4:35:40.954 $\pm$ 0.097 & 24:11:08.589 $\pm$ 0.108 &  4:35:40.975 $\pm$ 0.021 & 24:11:06.578 $\pm$ 0.024 &  2.828 $\pm$ 0.203 &  171.6 $\pm$   2.9 \\ 
IT Tau &  4:33:54.722 $\pm$ 0.020 & 26:13:27.201 $\pm$ 0.023 &  4:33:54.594 $\pm$ 0.030 & 26:13:25.494 $\pm$ 0.036 &  2.426 $\pm$ 0.056 &  225.3 $\pm$   0.9 \\ 
GK Tau &  4:33:34.572 $\pm$ 0.022 & 24:21:05.571 $\pm$ 0.025 &   ---   &   ---   &   ---   &   ---   \\ 
HN Tau &  4:33:39.376 $\pm$ 0.011 & 17:00:00.000 $\pm$ 0.012 &   ---   &   ---   &   ---   &   ---   \\ 
V710 Tau &  4:31:57.793 $\pm$ 0.008 & 18:21:37.655 $\pm$ 0.009 &   ---   &   ---   &   ---   &   ---   \\ 
IRAS 04298+2246 A &  4:32:49.120 $\pm$ 0.027 & 22:53:02.594 $\pm$ 0.033 &  4:32:49.232 $\pm$ 0.042 & 22:53:02.007 $\pm$ 0.050 &  1.653 $\pm$ 0.078 &  110.8 $\pm$   2.0 \\ 
IRAS 04298+2246 B &  4:32:49.433 $\pm$ 0.045 & 22:53:08.005 $\pm$ 0.054 &  4:32:49.433 $\pm$ 0.045 & 22:53:08.005 $\pm$ 0.054 &  0.525 $\pm$ 0.220 &  334.2 $\pm$  16.0 \\ 
HO Tau &  4:35:20.218 $\pm$ 0.009 & 22:32:14.312 $\pm$ 0.011 &   ---   &   ---   &   ---   &   ---   \\ 
UY Aur &  4:51:47.406 $\pm$ 0.010 & 30:47:13.234 $\pm$ 0.016 &  4:51:47.356 $\pm$ 0.027 & 30:47:12.693 $\pm$ 0.042 &  0.840 $\pm$ 0.054 &  230.0 $\pm$   2.7 \\ 
IRAS 05022+2527 &  5:05:22.824 $\pm$ 0.007 & 25:31:30.542 $\pm$ 0.009 &   ---   &   ---   &   ---   &   ---   \\ 
DS Tau &  4:47:48.609 $\pm$ 0.011 & 29:25:10.884 $\pm$ 0.016 &   ---   &   ---   &   ---   &   ---   \\ 
\enddata
\end{deluxetable*}

\subsection{Notes on individual sources}
\label{individ}

In this section, we discuss individual sources that were further examined or treated differently during
the analysis of the binary sample.  After our observations were obtained, we found that two objects
in our sample were known to be higher order multiples: IRAS 04298+2246 (JH 112) and
IRAS 04133+2758 (MHO 1/2), which are discussed further below.  We also reviewed
the evidence of youth for the undetected secondaries (GK Tau, HO Tau, DS Tau, V710 Tau, FX Tau, HN Tau and 
IRAS 05022+2527) and conclude that in three of these systems (GK Tau, HO Tau and DS Tau) the
observed companion is unlikely to be a young stellar object physically associated with the primary
star.  These three systems are not include in the
analysis of binary properties in \S \ref{binary}; further details are given below.

{\bfseries IRAS 04298+2246 (JH 112)}:  This source was confirmed as a binary by \citet{kra09} with a separation of 6\farcs6  but
later work \citep{kra11} identifies four components: Aa, Ab, Ba and Bb.
The ALMA observations detect all four components, resolving both IRAS 04298+2246 A and B into two sources with separations of 1\farcs7 and 0\farcs5 respectively.   For the B pair, the near-infrared position angle of 88\degr\ (White et al., in prep) indicates that
the secondary in this pair is brighter in the mm than the primary.
We treat this quadruple system as two binaries (A and B) in the analysis of binary systems. 
To assign stellar luminosities to the individual components, we use the A and B luminosities from \citet{and13} and
scale as $L_{\ast} \propto M_{\ast}^{1.75}$, using the mass ratio from \citet{kra11} for Aa and Ab and assuming an equal mass ratio for Ba and Bb.

{\bfseries IRAS 04113+2758 (MHO 1/2)}:  Following the WDS catalog component names and positions for the source WDS 04144+2806 AB \citep{mas01}, we associate IRAS 04113+2758 A with MHO 1 and IRAS 04113+2758 B with MHO 2, which is itself a close binary \citep[0\farcs05;][]{kra11}.  We detect both
of the widely spaced components, but we do not have the resolution to resolve the components of IRAS 04113+2758 B.    
Given the many close T Tauri binaries with substantial circumbinary disks,
e.g. GG Tau \citep{sim92}, UZ Tau E \citep{jen96b}, GW Ori \citep{mat95}, and DQ Tau \citep{mat97}, 
we include IRAS 04113+2758 in
the binary sample, even though it is a multiple, due to the very
small separation of the close pair.
In \S \ref{diskmass} the
adopted stellar mass is higher for IRAS 
04113+2758 B than for A, so we designate B as the primary star in this system.  With this assignment, the secondary
flux for this system is higher than the primary flux. However, we note that the stellar mass ratio is within 1$\sigma$ of unity and
the stellar mass estimation for B adopted from \citet{and13} treated B as a single star and did not derive separate stellar masses
for Ba and Bb.

{\bfseries GK Tau}: This source was first identified as a binary by \citet{rei93} where the companion is seen in CCD imaging; \citet{har94} provided colors for both sources.  Although \citet{duc99} measured an H$\alpha$ equivalent width of 45 \AA, they note that the spectrum has a poor signal-to-noise ratio.  Recently, Kraus (2014, in prep) identified GK Tau B as a background star.  We did not detect the secondary source and do not include this source in the analysis of binary systems.

{\bfseries HO Tau}: \citet{har94} were not successful in imaging the companion and conclude that the companion is likely to be a background star given its relative faintness and colors.  We did not detect the secondary source and do not include this source in the analysis of binary systems.

{\bfseries DS Tau}: \citet{mon91} did not detect H$\alpha$ from the companion and find that it is too faint to be a T~Tauri star in Taurus.  This is supported by the 2MASS colors, which are not red enough to be those of a young star (J-K=0.5 mag).  We did not detect the secondary source and do not include this source in the analysis of binary systems.

{\bfseries V710 Tau}: \citet{mcc06} detected both components in the mid-IR, the colors of both components are consistent with other T Tauri stars, and H$\alpha$ has been detected for both components \citep{whi01}. We did not detect the secondary source, but do include the mass limit in the analysis of binary systems.  
We use the \citet{kra11} mass ratio for this system.

{\bfseries FX Tau}: \citet{mcc06} detected both components in the mid-IR and the colors of both components are consistent with other T Tauri stars.  We did not detect the secondary source, but we do include the mass limit in the analysis of binary systems.

{\bfseries HN Tau}: \citet{woi01} measured resolved near-infrared photometry for both components, which are consistent with other T Tauri stars. 
 We did not detect the secondary source, but we do include the mass limit in the analysis of binary systems.

{\bfseries IRAS 05022+2527 (CIDA 9)}:  \citet{kra07} fit 2MASS images to obtain near-infrared magnitudes for the secondary component and the colors are consistent with a location in Taurus.   We did not detect the secondary source, but we do include the mass limit in the analysis of binary systems.

{\bfseries HBC 411 (CoKu Tau/3)}:  The position angle for the binary components at K band is 173\degr\ with a primary/secondary flux ratio of 3.9 \citep{whi01}.  In the millimeter, the secondary component is clearly brighter than the primary (Fig. \ref{fig:images}).  

\subsection{Stellar properties and disk mass}
\label{diskmass}
We adopt the stellar luminosities and masses from \citet{and13} for our sources, with the exception 
of V710 Tau B and IRAS 04298+2246 B as noted in the previous section; these stars were not included in \citet{and13}.
\citet{and13} derived luminosities by
assuming an effective temperature from the spectral type and fitting the SED for the stellar luminosity and
extinction, and derived masses by fitting to three different pre-main sequence stellar evolution grids.
Here we have selected their fits from the \citet{sie00} grids, as \citet{and13} found that these model masses were closest
to predicting the masses of those stars with dynamically determined masses.  The adopted stellar luminosities and
masses are given in Table \ref{tab:stellar}.  Assuming the dust is optically thin and isothermal, the conversion from flux ($F_{\nu}$) to
disk mass ($M_d$) is
\begin{equation}
M_d = \frac{F_{\nu}d^2}{\kappa_{\nu}X_gB_{\nu}(T_d)}.
\end{equation}
For comparison to the Taurus sample results of \citet{and13}, we use the same constants of
$d =$ 140 pc, dust-to-gas ratio $X_g$ = 0.01, dust opacity $\kappa_{\nu}$ = 2.3 cm$^2$g$^{-1}$ at 1.3 mm, and we use our
1.3 mm measured fluxes.  The uncertainty used for the flux measurement includes both the fit uncertainty
in Table \ref{tab:fit} and a 5\% absolute flux calibration uncertainty (ALMA memo 594).  For the
mean dust temperature $T_d$, we also adopt the \citet{and13} scaling of $T_d = 25 (L_{\ast}/L_{\odot})^{-1/4}$~K.  The
resulting disk masses are given in Table \ref{tab:stellar}.

\begin{deluxetable*}{llll}[h]
\tabletypesize{\scriptsize}
\tablewidth{0pt}
\
\tablecaption{Stellar properties and disk masses \label{tab:stellar}}
\tablehead{
\colhead{Source}     & \colhead{L$_{\ast}$} & \colhead{M$_{\ast}$} & \colhead{Disk mass} \\
& \colhead{(L$_{\odot}$)} & \colhead{(M$_{\odot}$)} & \colhead{(M$_{\odot}$)} \\
}
\startdata
FV Tau A & $    2.3 \pm   0.91 $ & $    1.2 ^{+  0.21} _{-  0.42} $ & $ 6.3\times10^{-4} \pm 1.4\times 10^{-4} $ \\ [1ex]
FV Tau B & $    1.4 \pm   0.57 $ & $   0.93 ^{+  0.19} _{-  0.17} $ & $ 7.2\times 10^{-4} \pm 1.6\times 10^{-4} $ \\  [1ex]
HBC 387 A & $   0.56 \pm   0.11 $ & $    0.3 ^{+  0.12} _{- 0.033} $ & $ 1.2\times 10^{-4} \pm 3.8\times 10^{-5} $ \\  [1ex]
HBC 387 B & $  0.064 \pm   0.04 $ & $   0.24 ^{+ 0.064} _{- 0.072} $ & $ 1.3\times 10^{-4} \pm 5.0\times 10^{-5} $ \\  [1ex]
FQ Tau A & $  0.086 \pm  0.031 $ & $   0.29 ^{+ 0.075} _{-  0.07} $ & $ 9.3\times 10^{-4} \pm 2.2\times 10^{-4} $ \\  [1ex]
FQ Tau B & $   0.12 \pm  0.043 $ & $   0.28 ^{+ 0.056} _{- 0.085} $ & $ 7.2\times 10^{-4} \pm 1.7\times 10^{-4} $ \\  [1ex]
FX Tau A & $   0.52 \pm   0.33 $ & $   0.48 ^{+  0.12} _{-  0.09} $ & $ 1.1\times 10^{-3} \pm 2.6\times 10^{-4} $ \\  [1ex]
FX Tau B & $   0.34 \pm   0.13 $ & $   0.24 ^{+ 0.055} _{- 0.049} $ & $ < 8.4\times 10^{-5} $ \\  [1ex]
 HK Tau A & $   0.44 \pm   0.15 $ & $   0.54 ^{+  0.17} _{-  0.11} $ & $ 5.8\times 10^{-3} \pm 1.3\times 10^{-3} $ \\  [1ex]
HK Tau B & $  0.027 \pm  0.015 $ & $   0.37 ^{+ 0.096} _{-  0.07} $ & $ 7.5\times 10^{-3} \pm 1.7\times 10^{-3} $ \\  [1ex]
DK Tau A & $    1.3 \pm   0.72 $ & $   0.71 ^{+  0.23} _{-  0.22} $ & $ 3.7\times 10^{-3} \pm 8.3\times 10^{-4} $ \\  [1ex]
DK Tau B & $   0.32 \pm   0.23 $ & $   0.47 ^{+  0.14} _{- 0.079} $ & $ 5.5\times 10^{-4} \pm 1.3\times 10^{-4} $ \\  [1ex]
IRAS 04113+2758 A & $    1.7 \pm     1.1 $ & $   0.35 ^{+ 0.082} _{- 0.066} $ & $ 2.4\times 10^{-2} \pm 5.5\times 10^{-3} $ \\  [1ex]
IRAS 04113+2758 B & $    1.4 \pm    1.2 $ & $   0.48 ^{+  0.21} _{- 0.062} $ & $ 1.6\times 10^{-2} \pm 3.6\times 10^{-3} $ \\  [1ex]
UY Aur A & $      1 \pm   0.45 $ & $   0.59 ^{+  0.19} _{-  0.18} $ & $ 2.7\times 10^{-3} \pm 6.2\times 10^{-4} $ \\  [1ex]
UY Aur B & $   0.52 \pm   0.23 $ & $   0.32 ^{+ 0.093} _{- 0.048} $ & $ 1.3\times 10^{-3} \pm 2.9\times 10^{-4} $ \\  [1ex]
IRAS 05022+2527 A & $  0.098 \pm  0.078 $ & $   0.62 ^{+ 0.075} _{-  0.17} $ & $ 1.0\times 10^{-2} \pm 2.3\times 10^{-3} $ \\  [1ex]
IRAS 05022+2527 B & $  0.082 \pm   0.13 $ & $   0.39 ^{+  0.15} _{-  0.16} $ & $ < 1.7\times 10^{-4} $ \\  [1ex]
HBC 411 A & $    0.6 \pm   0.16 $ & $   0.48 ^{+  0.12} _{-  0.09} $ & $ 2.9\times 10^{-4} \pm 7.9\times 10^{-5} $ \\  [1ex]
HBC 411 B & $    0.2 \pm  0.074 $ & $   0.16 ^{+  0.15} _{- 0.042} $ & $ 1.3\times 10^{-3} \pm 3.0\times 10^{-4} $ \\  [1ex]
IT Tau A & $    1.4 \pm   0.49 $ & $    1.4 ^{+  0.17} _{-  0.18} $ & $ 8.3\times 10^{-4} \pm 1.9\times 10^{-4} $ \\  [1ex]
IT Tau B & $   0.21 \pm  0.096 $ & $   0.23 ^{+ 0.061} _{- 0.065} $ & $ 9.2\times 10^{-4} \pm 2.1\times 10^{-4} $ \\  [1ex]
HN Tau A & $   0.42 \pm   0.55 $ & $   0.91 ^{+  0.21} _{-  0.19} $ & $ 2.4\times 10^{-3} \pm 5.3\times 10^{-4} $ \\  [1ex]
HN Tau B & $  0.028 \pm  0.019 $ & $   0.19 ^{+ 0.055} _{- 0.062} $ & $ < 2.8\times 10^{-4} $ \\  [1ex]
V710 Tau A & $   0.57 \pm   0.19 $ & $   0.57 ^{+  0.15} _{-  0.12} $ & $ 9.3\times 10^{-3} \pm 2.1\times 10^{-3} $ \\  [1ex]
V710 Tau B & $   0.47 \pm   0.24 $ & $    0.4 ^{+   0.1} _{- 0.084} $ & $ < 1.1\times 10^{-4} $ \\  [1ex]
IRAS 04298+2246 Aa & $    1.1 \pm   0.39 $ & $   0.95 ^{+  0.14} _{-   0.2} $ & $ 4.5\times 10^{-4} \pm 1.1\times 10^{-4} $ \\  [1ex]
IRAS 04298+2246 Ab & $ 0.0012 \pm 0.00041 $ & $  0.016 ^{+ 0.016} _{-0.0086} $ & $ 7.2\times 10^{-3} \pm 1.7\times 10^{-3} $ \\  [1ex]
IRAS 04298+2246 Ba & $ 0.0025 \pm 0.0017 $ & $   0.13 ^{+ 0.033} _{- 0.026} $ & $ 3.7\times 10^{-4} \pm 1.4\times 10^{-4} $ \\  [1ex]
IRAS 04298+2246 Bb & $ 0.0025 \pm 0.0017 $ & $   0.13 ^{+ 0.033} _{- 0.026} $ & $ 3.4\times 10^{-3} \pm 8.4\times 10^{-4} $ \\  [1ex]
\enddata
\end{deluxetable*}

\begin{figure*}[h]
\includegraphics[angle=270,width=6.5in]{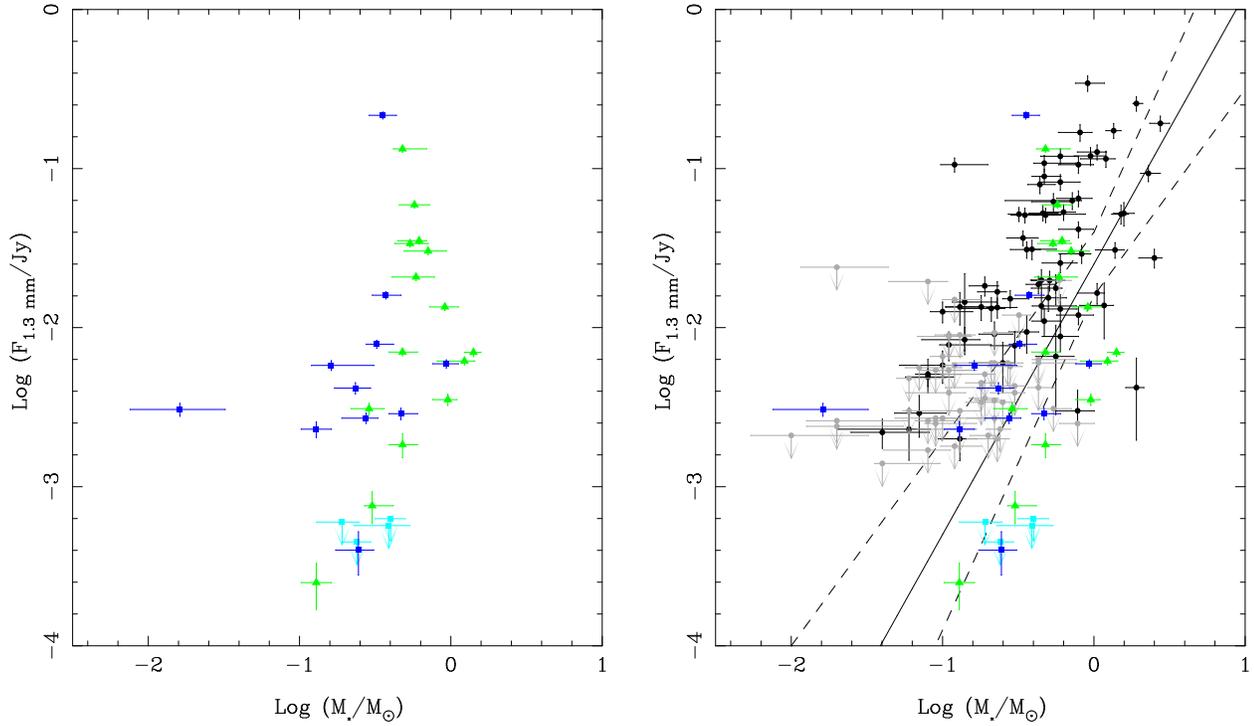}
\caption{Left) 1.3 mm flux for our primaries (green triangles) and secondaries (squares: dark blue - detections; light blue - non-detections).  Right) Our data (same symbols as left panel) compared to the Taurus sample single stars from \citet{and13} (black - detections; grey - non-detections) as a function of stellar mass, using the \citet{sie00} model fits in \citet{and13}.  The solid and dashed lines show the linear best fit and 95\% confidence boundaries from \citet{and13} to the complete Taurus sample.
\label{fig:flux}
}
\end{figure*}

\begin{figure*}[h]
\includegraphics[angle=270,width=6.5in]{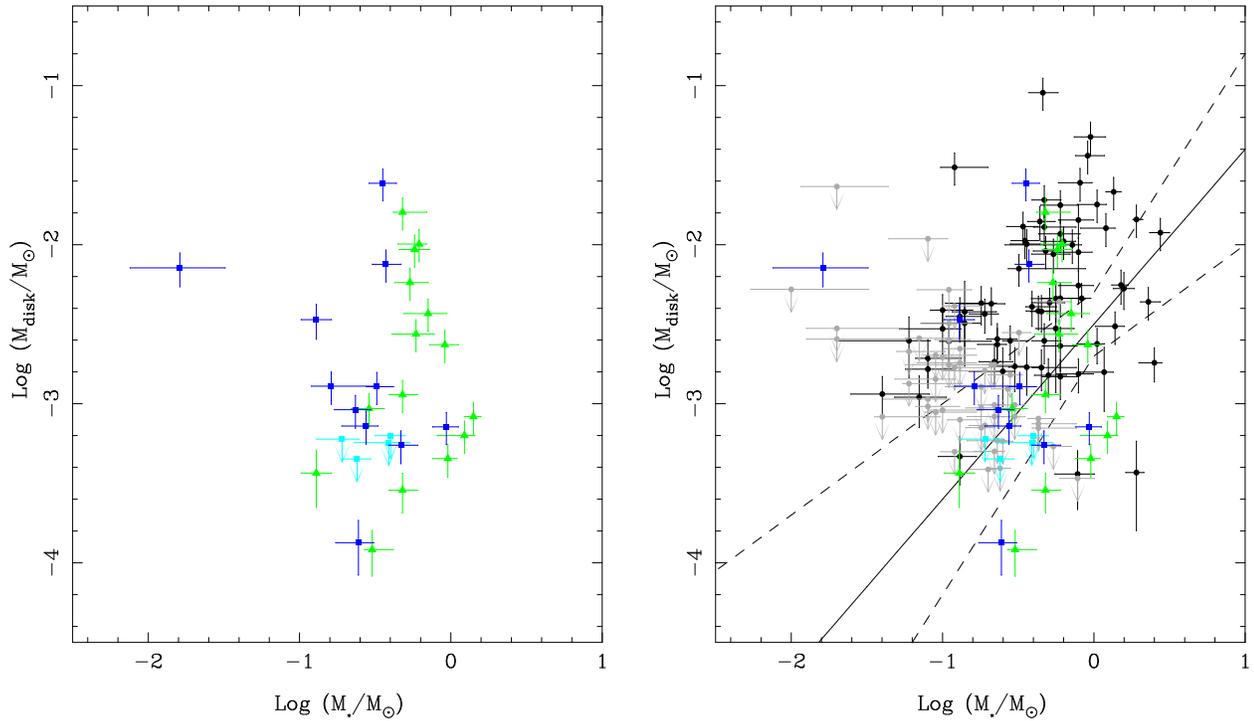}
\caption{As for Fig. \ref{fig:flux} but with the disk mass as a function of stellar mass.
\label{fig:diskm}
}
\end{figure*}

To probe some of the issues of disk formation and evolution in binary systems raised in \S \ref{intro}, we first compared the
fluxes and disk masses of the primaries to those of the secondaries.  As can
be seen in the left-hand panels of Figures \ref{fig:flux}
and \ref{fig:diskm}, the spread in flux and disk mass is larger than the spread in stellar mass.
To assemble a set of comparison single stars in Taurus, we used the sample in \citet[][their Table 2]{and13}, and removed all known multiple sources using the list in \citet{kra11}.  As pointed out by \citet{and13}, there are likely to be some sources labeled as single that in fact have close companions, as multiplicity surveys are not complete at the lowest stellar masses and closest separations, but we do not attempt to correct for this.  
The right panels of Fig. \ref{fig:flux} and \ref{fig:diskm} compare our measured 1.3 mm fluxes and derived disk masses to
the sample of single Taurus stars.  
The higher sensitivity of our observations reveals several detections and upper limits
significantly below the limits of previous work, suggesting that a population of lower mass disks remains to
be detected, particularly around the lower mass stellar hosts.

To quantitatively compare our sample to previous work, we used the survival analysis methods
described in \citet{fei85} and \citet{iso86} as implemented in the ASURV package \citep{lav92}
to calculate the correlation probabilities in the presence of upper limits.   
As recommended in \citet{fei85}, we ran multiple versions of the relevant
tests to compare the measurements.
When comparing the measured flux, we used the univariate two-sample test methods 
Gehan, Peto and Peto, and Peto and Prentice from ASURV \citep{lav92}.  These tests show that the probability
of the primary and the single stars coming from the same population is 20-60\% while
the probability that the secondaries and singles are drawn from the same population
is low (8-17\%).  The strongest result was for the test that the primaries and secondaries 
came from the same population, which has a probability of only 3-4\%.   However, given
the limited sample size, the comparison of the primaries and secondaries may be significantly biased
by the lower stellar mass for the secondaries.  These results are similar to those
found by \citet{har12} in comparison of fluxes from singles, primaries, and secondaries.
We also calculated the linear regression between the stellar mass and the disk
mass for the primaries and secondaries as separated samples using ASURV \citep{lav92}.  The resulting
fits (using both the parametric EM algorithm and the Buckley-James method)
are not well constrained, but agree within the uncertainties with the slope
of log disk mass to log stellar mass found for all Taurus disks by \citet{and13}.

\begin{figure}
\begin{center}
\includegraphics[width=3.5in]{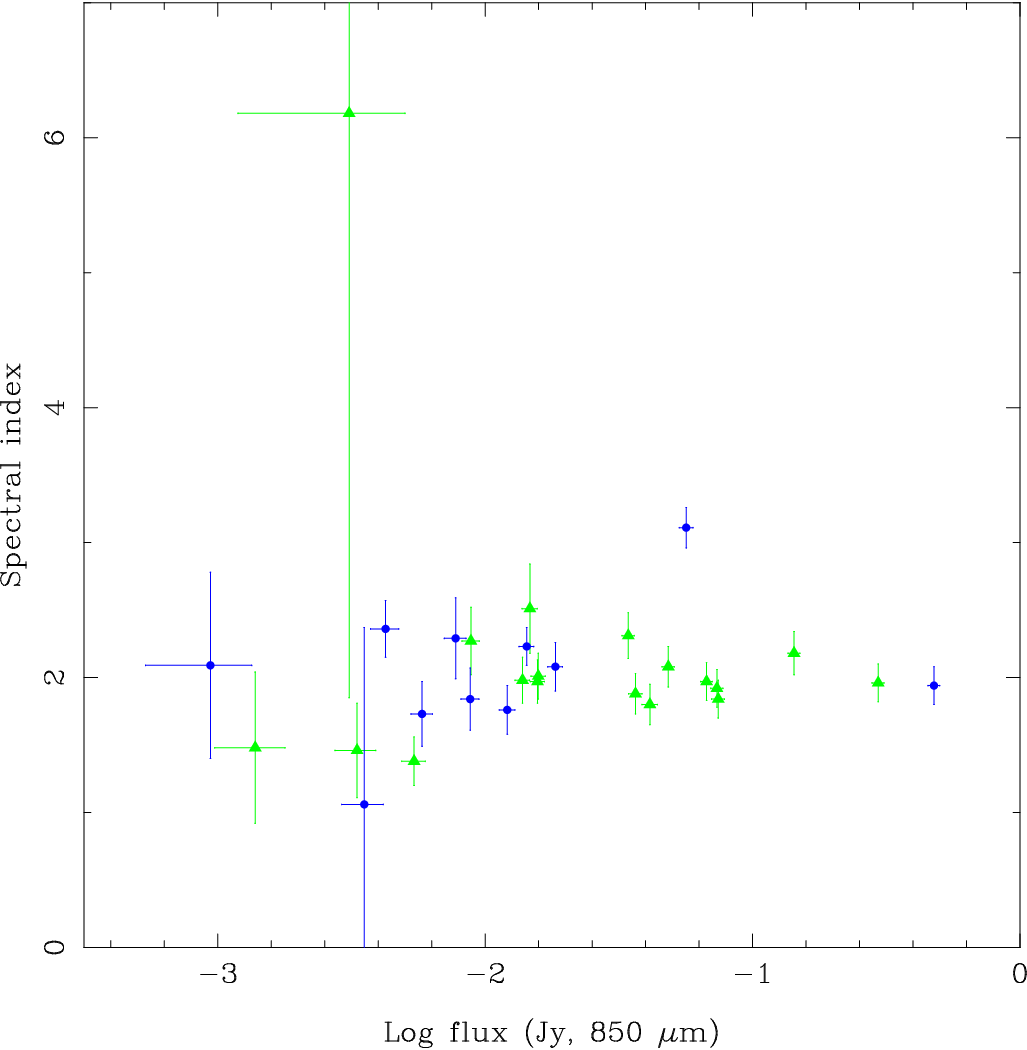}
\caption{The 850 $\mu$m/1.3 mm spectral index as a function of the 850 $\mu$m flux for all detected components, with primaries plotted as green triangles and secondaries as blue circles.
\label{fig:specindex}
}
\end{center}
\end{figure}

As all detected sources were detected in both bands, we calculated the
spectral index $\alpha$ between 1.3 mm and 850 $\mu$m, where $F_{\rm 850 \mu m}/F_{\rm 1.3 mm}= (\lambda_{\rm 850 \mu m}/\lambda_{1.3 mm})^{\alpha}$ 
(Fig. \ref{fig:specindex}).  The spectral index values calculated for both the primaries and
secondaries have an average of 2.1 and are similar to previous surveys of T Tauri stars \citep{and05}.

\subsection{Binary comparisons}
\label{binary}

\begin{figure*}[h]
\includegraphics[angle=270,width=6.5in]{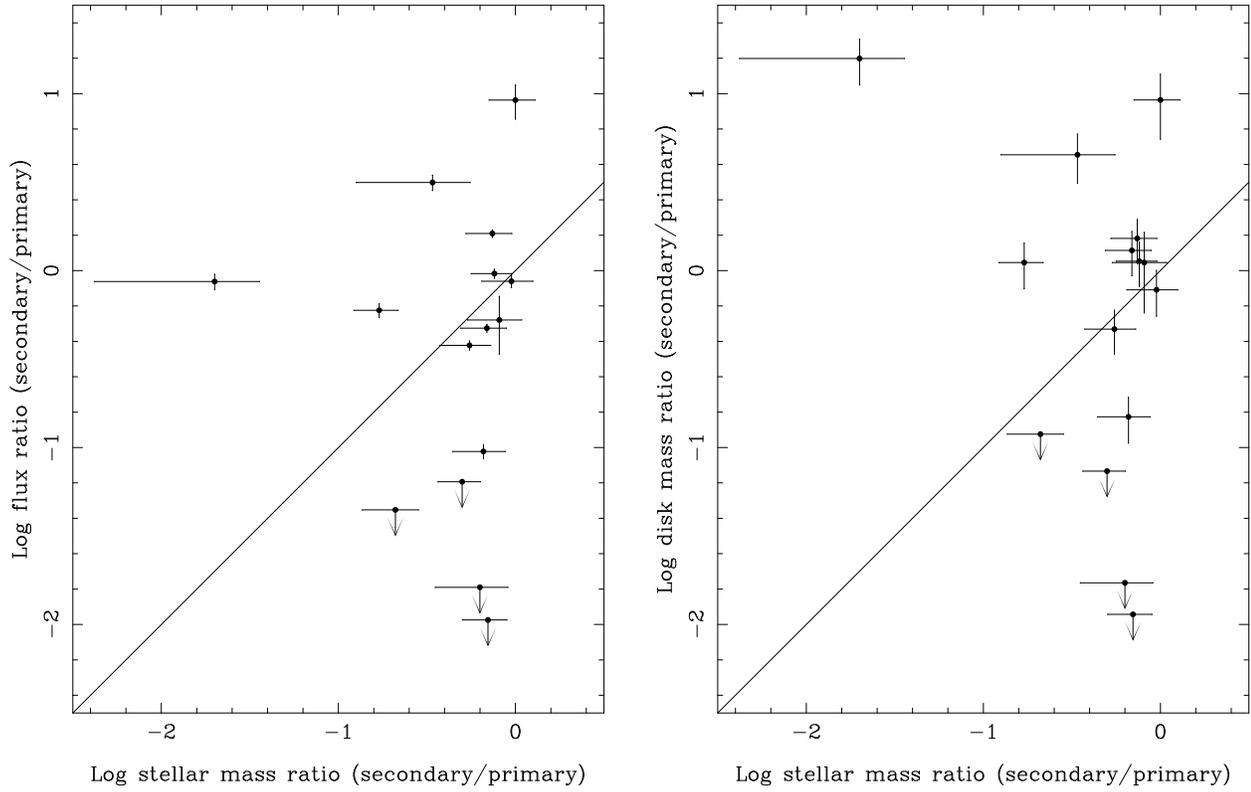}
\caption{The 1.3 mm flux ratio (left) and calculated disk mass ratio (right) compared to the stellar mass ratio for each binary system.  A line tracing equal ratios is shown for comparison.  Strikingly, there appears to be no relationship between the stellar mass ratio and the flux or disk mass ratio.
\label{fig:ratios}
}
\end{figure*}

\begin{figure*}[h]
\includegraphics[angle=270,width=6.5in]{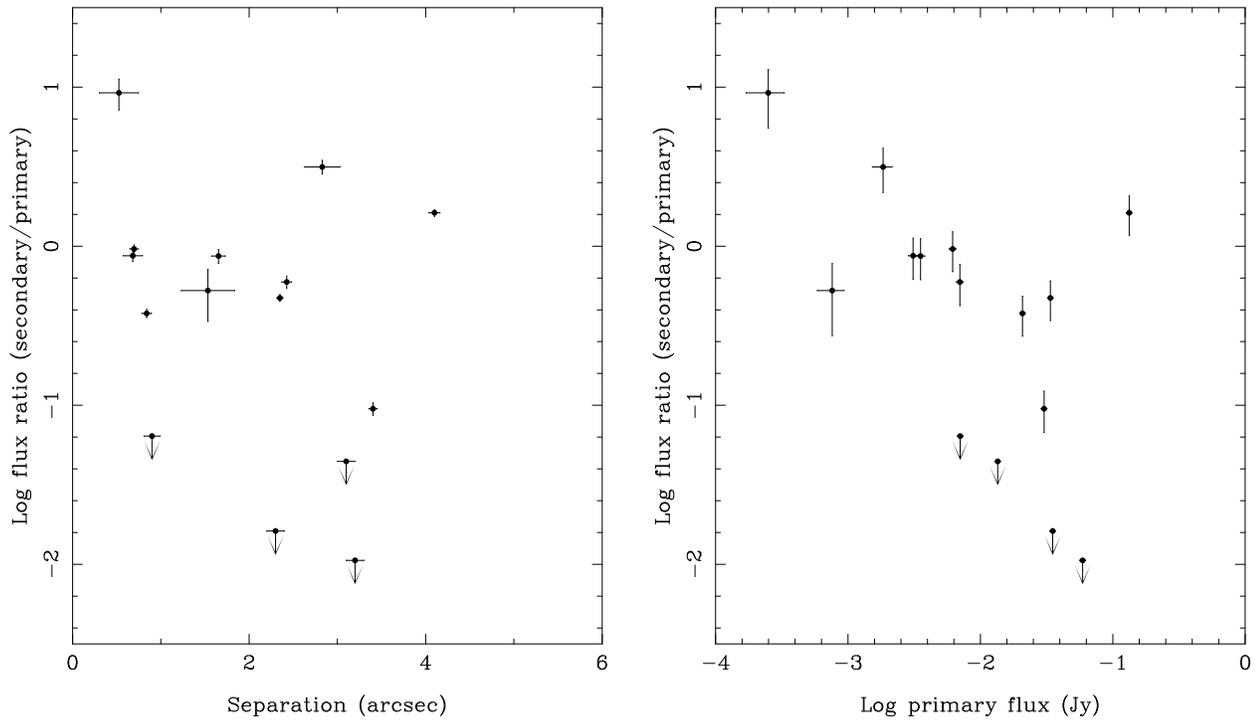}
\caption{The 1.3 mm flux ratio as a function of the binary separation (left) and primary flux (right).
\label{fig:ratios2}
}
\end{figure*}

We have also compared the disks within each binary system for the 15 systems where
we are confident of the pre-main sequence status of each component.   In Fig. \ref{fig:ratios}, we plot
the secondary/primary flux and disk mass ratios as a function of the secondary/primary
stellar mass ratio and in Fig. \ref{fig:ratios2}, we plot the flux ratio
as a function of projected separation and primary flux.
The previous surveys by \citet{jen03} and \citet{har12} both found that
if two components were detected, the primary always had the higher flux.
In this larger and more sensitive sample, we always detect the primary disk, but 
we find two systems, IRAS 04298+2246 B and
HBC 411, where the secondary flux is significantly (i.e. $> 1\sigma$) higher (discounting
IRAS 04113+2758 given the issues discussed in \S \ref{individ}).  
We note that we have assumed an equal stellar mass ratio for the components of IRAS 04298+2246 B,
where the primary status is assigned on the basis of the near-infrared flux ratio.
If the primary has a higher stellar mass than the secondary,
then the stellar mass ratio will be less than one and this system will be even more discrepant.
In the comparison of the disk mass,
where the stellar luminosity is factored in via the derived dust temperature (\S  
\ref{diskmass}), a third system, IRAS 04298+2246 A, also has a significantly higher
secondary disk mass.  

We can use these results to examine predictions for binary disk masses.
The observational result that the primary always has a disk if the
system is detected is consistent with the predictions of binary formation models
by \citet{bat97} and \citet{bat00}.     A second test of these
models is the ratio of the disk masses within the binary system.
\citet{bat00}
find that the primary disk should be more massive unless
the circumprimary disk accretes on a shorter time-scale or the
ongoing accretion is due to material from a circumbinary disk; however,
\citet{och05} use a different numerical viscosity and find that the primary
accretion from the circumbinary material is always higher than the secondary rate.   Although the
ALMA observations were not designed for sensitivity to extended emission,
there is no evidence for substantial circumbinary emission in any of these systems.
In the extended configuration used for these observations, the maximum
scale for extended emission to be detected was 3\farcs0 at 850 $\mu$m and 4\farcs4 at 1.3 mm.
While no circumbinary emission is detected, this material may have dissipated more
quickly than the circumstellar disks, so we cannot constrain whether there could be disk mass differences
due to differential accretion from a circumbinary envelope.

While the binary disk formation models of \citet{bat97}, \citet{bat00}, and  \citet{och05} address the formation of circumprimary and circumsecondary disks, they do not cover the subsequent viscous evolution and dissipation in the disk, which may be a factor for our sample as the mean age for Class II sources in Taurus was estimated to be 2.5-3 Myr by \citet{and13}.  Disk evolution has been shown to be dependent  on stellar mass in other star formation regions.  Two studies of clusters aged $\sim$5 Myr showed similar results; in Upper Sco, \citet{car06} found 20\% of K and M stars retained their disks while none of the F and G stars did, and in NGC 2362, \citet{dah07} found a disk fraction of 19\% for stellar masses less than 1.2 M${_\odot}$, while none of the more massive stars still had a disk.  While the stars in our sample range from 0.02 to 1.4 M${_\odot}$, with only IT Tau A above the 1.2 M${_\odot}$ cutoff found in the studies of older clusters, the general
trend of faster dissipation for higher mass stars would affect our comparison of primary
and secondary disk masses, as by definition the
primary always has a higher stellar mass.
To disentangle the separate roles of stellar mass and multiplicity in the evolution of the disk mass, a comparison
of the single, primary, and secondary disk masses as a function of
stellar mass is needed.  As Fig. \ref{fig:diskm} shows, most of the stars with masses less than
0.6 M$_{\odot}$ remain undetected in Taurus and more observations are needed
before these dependencies can be quantified.

As Fig. \ref{fig:diskm} demonstrates, at a given stellar mass, there is a large scatter in disk mass; \citet{and13} measured a standard deviation of 0.7 dex in the log disk mass around the best linear fit as a function of stellar mass. 
If this scatter is due to initial conditions and/or disk evolution factors that vary on scales greater than
the binary systems, i.e. are similar for a given primary and secondary, 
we would expect the disk mass ratios to be correlated with the stellar mass ratios.  Using the ASURV survival analysis code \citep{lav92} the probability that the log of the stellar mass ratio and the log of the disk mass ratio are correlated is 8-65\% (for the Cox proportional hazard and Kendall's tau tests), while \citet{and13} found a correlation in log stellar mass/log disk mass of $>99.9\%$ for their entire sample of Taurus objects.  This suggests that the factors which determined the disk masses for these
binaries, both initial conditions and disk evolution, are not constant between the components.
As discussed above, there is a known impact of stellar mass on the disk evolution and if this
could be quantified, these binary systems could then be used as probe of other effects
such as initial conditions and dynamical interactions.

\section{Conclusions}
\label{conclusions}

We obtained ALMA observations of 17 young stellar multiples in Taurus and the sensitivity achieved resulted in several new detections of disks.  Two of these new detections are of primaries and six are secondaries.  Disks around all of the primary stars are detected, but four of the secondaries that show other signatures of youth and inner disk accretion (\S \ref{individ}) remain undetected and are shown to have disk masses less than $10^{-4}$~M$_{\odot}$ for standard disk parameters.  The new ALMA detections are generally at flux levels less than 5 mJy at 1.3 mm, below the limits of previous surveys, suggesting that many undetected objects in Taurus may simply have disk masses below $10^{-3}$~M$_{\odot}$ and a more sensitive, systematic survey of the Taurus population is needed to further quantify the stellar to disk mass relation seen by \citet{and13} for the lowest mass stars.  We also examined the properties of the binary systems and
have the following conclusions:

\begin{itemize}
\item The majority of our new detections were for secondary sources and
for these wide binaries ($> 100$ AU), the secondary disk
fraction is somewhat higher than shown in previous studies.  We 
found 11 of 
the 15 bona fide young stellar binaries have disks with masses  $\geq 10^{-4}$M$_{\odot}$ around both stars,
while \citet{jen03} detected a circumsecondary disk in 1 of 4 systems and \citet{har12}
detected a circumsecondary disk in 6 of 12 systems where the components were resolved. 
There is significant overlap in the samples between these studies and
the new detections are primarily due to the higher sensitivity of the ALMA observations.
The newly detected primary disk masses and most of the secondary disk masses are
considerably smaller than the minimum mass solar nebula, but this is not surprising
given that the host stars are
generally less than 1 M$_{\odot}$.  While it may be
difficult for massive planets to form in these less massive disks, models of
core accretion around lower-mass stars show that they may be able to
form cores for lower-mass planets \citep[e.g.][]{lau04} and if the
disk mass has evolved, larger planets may
have formed earlier when the disk was more massive.

\item
In two binary systems the secondary disk has a higher mm flux than the primary
disk.  This has not been seen in previous, smaller surveys and is counter to
predictions of formation models where the infalling material is directly accreted onto
the primary or secondary disk as opposed to accreting onto a circumbinary structure.
This result could be explained by faster dissipation of the primary
disk, which has been shown to be a function of stellar mass. 

\item For this sample of wide binaries,  the secondary/primary
disk mass ratio is not correlated with the secondary/primary stellar
mass ratio.  This suggests that for these binary systems, any
environmental factors shared between the two components
that could affect the initial disk mass and disk evolution
are not the dominant factor in determining the range of disk masses for
a given stellar mass.  

\end{itemize}

From these conclusions, it is clear that binaries
do not follow a simple pattern of primary/secondary disk mass distribution; 
therefore, care should be taken when assigning
flux to components in unresolved systems.

\acknowledgments

This paper makes use of the following ALMA data: ADS/JAO.ALMA\#2011.0.00150.S. ALMA is a partnership of ESO (representing its member states), NSF (USA) and NINS (Japan), together with NRC (Canada) and NSC and ASIAA (Taiwan), in cooperation with the Republic of Chile. The Joint ALMA Observatory is operated by ESO, AUI/NRAO and NAOJ.  The National Radio Astronomy Observatory is a facility of the National Science Foundation operated under cooperative agreement by Associated Universities, Inc.

We thank Scott Schnee at NRAO for extensive help in reducing the ALMA data and Sean Andrews,
Adam Kraus, and Russel White for useful discussions.  This paper makes use of the
ASURV package, Rev 1.2 \citep{lav92}.  

Facilities: \facility{ALMA}


\end{document}